\documentclass[pra,twocolumn]{revtex4}
\newcount\tipo 
\font\titolone=cmbx10 scaled\magstep 2%
\font\sc=cmcsc10%
\font\ottorm=cmr8%
\def\st{\scriptstyle}%
\font\tenmib=cmmib10 \font\eightmib=cmmib8
\font\sevenmib=cmmib7\font\fivemib=cmmib5 
\font\ottoit=cmti8\font\fiveit=cmti5\font\sixit=cmti6%%
\font\fivei=cmmi5\font\sixi=cmmi6\font\ottoi=cmmi8
\font\ottorm=cmr8
\font\ottosy=cmsy8\font\sixsy=cmsy6\font\fivesy=cmsy5%%
\font\ottobf=cmbx8\font\sixbf=cmbx6\font\fivebf=cmbx5%
\font\ottocss=cmcsc8%
\def\ottopunti{\def\rm{\fam0\ottorm}\def\it{\fam6\ottoit}%
\def\bf{\fam7\ottobf}%
\textfont1=\ottoi\scriptfont1=\sixi\scriptscriptfont1=\fivei%
\textfont2=\ottosy\scriptfont2=\sixsy\scriptscriptfont2=\fivesy%
\textfont4=\ottocss\scriptfont4=\sc\scriptscriptfont4=\sc%
\textfont5=\eightmib\scriptfont5=\sevenmib\scriptscriptfont5=\fivemib%
\textfont6=\ottoit\scriptfont6=\sixit\scriptscriptfont6=\fiveit%
\textfont7=\ottobf\scriptfont7=\sixbf\scriptscriptfont7=\fivebf%
\setbox\strutbox=\hbox{\vrule height7pt depth2pt width0pt}%
\normalbaselineskip=9pt\rm}
\let\nota=\ottopunti%
\mathchardef\Ba   = "050B  %alfa
\mathchardef\Bb   = "050C  %beta
\mathchardef\Bg   = "050D  %gamma
\mathchardef\Bd   = "050E  %delta
\mathchardef\Be   = "0522  %varepsilon
\mathchardef\Bee  = "050F  %epsilon
\mathchardef\Bz   = "0510  %zeta
\mathchardef\Bh   = "0511  %eta
\mathchardef\Bthh = "0512  %teta
\mathchardef\Bth  = "0523  %varteta
\mathchardef\Bi   = "0513  %iota
\mathchardef\Bk   = "0514  %kappa
\mathchardef\Bl   = "0515  %lambda
\mathchardef\Bm   = "0516  %mu
\mathchardef\Bn   = "0517  %nu
\mathchardef\Bx   = "0518  %xi
\mathchardef\Bom  = "0530  %omi
\mathchardef\Bp   = "0519  %pi
\mathchardef\Br   = "0525  %ro
\mathchardef\Bro  = "051A  %varrho
\mathchardef\Bs   = "051B  %sigma
\mathchardef\Bsi  = "0526  %varsigma
\mathchardef\Bt   = "051C  %tau
\mathchardef\Bu   = "051D  %upsilon
\mathchardef\Bf   = "0527  %phi
\mathchardef\Bff  = "051E  %varphi
\mathchardef\Bch  = "051F  %chi
\mathchardef\Bps  = "0520  %psi
\mathchardef\Bo   = "0521  %omega
\mathchardef\Bome = "0524  %varomega
\mathchardef\BG   = "0500  %Gamma
\mathchardef\BD   = "0501  %Delta
\mathchardef\BTh  = "0502  %Theta
\mathchardef\BL   = "0503  %Lambda
\mathchardef\BX   = "0504  %Xi
\mathchardef\BP   = "0505  %Pi
\mathchardef\BS   = "0506  %Sigma
\mathchardef\BU   = "0507  %Upsilon
\mathchardef\BF   = "0508  %Fi
\mathchardef\BPs  = "0509  %Psi
\mathchardef\BO   = "050A  %Omega
\mathchardef\BDpr = "0540  %Dpr
\mathchardef\Bstl = "053F  %*
\def\fiat{}
\let\a=\alpha \let\b=\beta    \let\d=\delta  \let\e=\varepsilon
\let\z=\zeta        
\let\m=\mu    \let\n=\nu         \let\p=\pi     \let\r=\rho
\let\s=\sigma \let\t=\tau   \let\f=\varphi 
      
 \let\D=\Delta   
        \let\Ps=\Psi
\let\O=\Omega 

\def\CC{{\cal C}}

\let\ig=\int
\let\io=\infty
\let\==\equiv
\let\0=\noindent
\let\Dpr=\BDpr

\def\\{\hfill\break}

\def\*{\vskip3mm} 
\let\dpr=\partial
\def\defi{\,{\buildrel def\over=}\,}
\def\V#1{{\underline#1}}
\def\media#1{{\langle#1\rangle}}
\def\fra#1#2{{#1\over#2}}
\def\crcl{\,\raise.5mm\hbox{$\st\rm o$}\,}%
\def\otto{\,{\kern-1.truept\leftarrow\kern-5.truept\to\kern-1.truept}\,}
\def\tende#1{\,\vtop{\ialign{##\crcr\rightarrowfill\crcr
 \noalign{\kern-1pt\nointerlineskip} \hskip3.pt${\scriptstyle
 #1}$\hskip3.pt\crcr}}\,}

\newbox\strutboxa
\setbox\strutboxa=\hbox{\vrule height8.5pt depth2.25pt width0pt}
\def\struta{\relax\ifmmode\copy\strutboxa\else\unhcopy\strutboxa\fi}
\def\W#1{#1_{\kern-3pt\lower7.5pt\hbox{$\widetilde{}$}}\kern2pt\,\struta}
\def\T#1{{#1_{\kern-3pt\lower7pt\hbox{$\widetilde{}$}}\kern3pt}}
\def\VV#1{{\underline #1}_{\kern-3pt
\lower7pt\hbox{$\widetilde{}$}}\kern3pt\,}

\newdimen\xshift \newdimen\xwidth \newdimen\yshift \newdimen\ywidth

\def\ins#1#2#3{\vbox to0pt{\kern-#2\hbox{\kern#1 #3}\vss}\nointerlineskip}

\def\eqfig#1#2#3#4#5{%
\par\xwidth=#1\xshift=\hsize\advance\xshift%
by-\xwidth\divide\xshift by 2%
\yshift=#2\divide\yshift by 2%
{\hglue\xshift \vbox to #2{\vfil%
#3\includegraphics{#4.eps}%
}\hfill\raise\yshift\hbox{#5}}}%

\newcommand\revtex{{R\kern-0.4mm\lower0.5mm\hbox{E}\kern-0.4mm V\kern-0.3mm%
\lower0.5mm\hbox{T}\kern-0.4mm E\kern-.3mm \lower0.5mm\hbox{X}}}

\fiat
\begin{document}

\centerline{\titolone Quantum Nonequilibrium}
\centerline{\titolone and Entropy Creation}
\*
\centerline{\bf Giovanni Gallavotti}

\centerline{Fisica and I.N.F.N. Roma 1}
\centerline{13 Jan 2007
%\today
}
\*

\0{\bf Abstract: \it In sharp contrast to the corresponding classical
  systems cases it is not yet understood how to define a mechanical
  quantity with the interpretation of entropy creation rate for
  nonequilibrum stationary states of finite quantum systems with
  finite thermostats. Some aspects of this problem are discussed here
  in cases in which identifying entropy creation rate as a mechanical
  observable might be possible.}

\*
%\0{PACS: 47.52, 05.45, 47.70, 05.70.L, 05.20, 03.20}
\0{\nota\sl Keywords: \rm Quantum Nonequilibrium, Chaotic Hypothesis,
  Fluctuation Theorem, Entropy, Large Deviations, Nonequilibrium
  Statstical Mechanics\vfil}
\*
\0{\bf 1. Classical systems and thermostats}
%\section{Classical systems and thermostats}
\setcounter{section}{1}\setcounter{equation}{0}
\*

The aim of this paper is to propose a notion of thermostat acting {\it
  reversibly} on a quantum mechanical system which is analogous to the
  Gaussian or Nos\'e-Hoover thermostats that have been so important in
  recent times for the development of research non nonequilibrium. At
  the same time we remark that ``entropy creation'' can be associated
  with the thermostats discussed here and can be given a meaning that
  makes it susceptible of experimental measurements. This is important
  also because it allow us to extend the ``chaotic hypothesis'' to
  such systems and to test the ``fluctuation relation'' that holds, at
  least as a formal consequence. I begin by recalling the notion of
  classical thermostat that will be genreralized here and some results
  concerning classical nonequilibrium statistical mechanics.

A {\it thermostat} is, physically, a device that extracts heat created
inside a mechanical system subject to non conservative forces, thus
allowing control of the energy build-up: in this way a forced system
can reach a {\it stationary state}. Such a state is however deeply
different from an equilibrium state. Heat, matter or electric currents
may be present and the dissipation associated with the thermostats
implies that the probability distribution $\m$ that describes the
statistical properties of the system is not even close to the familiar
Maxwell-Boltzmann distributions for equilibrium. This feature makes
the problem of the statistics of stationary nonequilibria particularly
interesting and challenging.

In the classical case a rather general model for a system
in contact with thermostats is represented in Fig.1, \cite{Ga06c}.
\eqfig{240pt}{90pt}{
\ins{90pt}{60pt}{$\V X_0,\V X_1,\ldots,\V X_n$}
\ins{60pt}{27pt}{$\ottoi\st\ddot{\V X}_{0i}=-\dpr_i U_0(\V X_0)-\sum_{j}
\dpr_i U_j(\V X_0,\V X_j)+\V F_i$}
\ins{60pt}{10pt}{$\ottoi\st\ddot{\V X}_{ji}=-\dpr_i U_j(\V X_j)-
\dpr_i U_j(\V X_0,\V X_j)-\a_j \V{{\dot X}}_{ji}$}
}{fig0}{}
\\
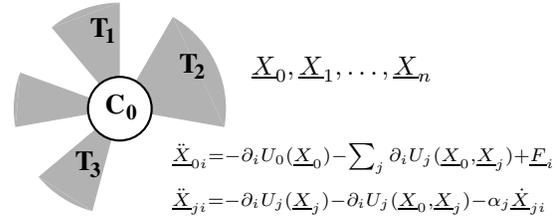
\0{Fig.1\ottorm Reservoirs occupy finite regions outside $\CC_0$, {\it
e.g.} sectors $\CC_j\subset R^3$, $j=1,2\ldots$. Their particles (with
mass $1$) are constrained to have a {\it total} kinetic energy $K_j$
constant, by suitable forces, so that the reservoirs ``temperatures''
$T_i$ are well defined. Fixed $j$ the label $i$ is in
$1,\ldots.N_j$.\vfil}
%%Note that the figure contains tex text superposed

The equations of motion for the particles positions $\V X_0,\V
X_1,\ldots,\V X_n$ are written in Fig.1 in terms of 
\*

\0(1) The potential energies for the $N_j$ particles inside each
region $\CC_j$,
$U_j(\V X_j)$, which will be assumed bounded, for simplicity; all
masses are $m=1$, also for simplicity,
\\
\0(2) The potential energies between particles in $\CC_0$ and in
$\CC_j$:
$U_j(\V X_0,\V X_j)$, also assumed bounded,
\\
\0(3) The external, non conservative, positional forces $\V F$ acting 
on the particles in $\CC_0$,
\\
\0(4) The thermostat forces $\a_i \V{{\dot X}}_{i}$ which are so defined
that the total kinetic energy $K_j=\fra12 \V{{\dot X}}_j^2$ in each
thermostat is strictly constant. Such forces will be imagined realized 
by imposing constancy of $K_j$ via {\it Gauss' principle}, 
\cite[Sec.9]{Ga00}. This gives, \cite{Ga06c},

\begin{equation}
\a_j\=\fra{W_j-\dot U_j}{2 K_j}\label{e1.1}\end{equation}
where $W_j$ is the work done per unit time by the system particles on
the $j$--th thermostat particles and $U_j$ is the internal potential
energy of the $j$--th reservoir: $W_j = -\V{{\dot X}}_j\cdot\Dpr_{\V
X_j}U_j(\V X_0,\V X_j)$. {\it Thus $W_j$ will be identified with the 
heat $Q_j$ ceded per unit time by the system to the $j$-th
thermostat.} 
\\ 
\0(5) The value of the constant $K_j$ will be written
$K_j=\fra32 N_j k_B T_j$ ($k_B=$ Boltzmann's constant) and will define
the {\it temperature} of the $j$-th thermostat, \cite{Ga06c}.  \*

A brief computation yields the divergence of the
equations of motion, {\it i.e.} of the phase space contraction rate,
for the velocity--position coordinates $\V{{\dot X}},\V X$

\begin{equation}\s(\V{{\dot X}},\V X)=\,\e(\V{{\dot X}},\V X)+\dot R(\V X)
\label{e1.2}\end{equation}
where, remarkably, $\e(\V{{\dot X}},\V X)$ can be interpreted as the {\it entropy
creation rate}

\begin{equation}\label{e1.3}
\e(\V{{\dot X}}, \V X)=\sum_{j>0} \fra{Q_j}{k_B T_j},\quad R(\V X)=
\sum_{j>0} \fra{U_j}{k_B T_j},
\end{equation}
Eq.(\ref{e1.3}) are correct up to $O(N^{-1})$ if $N=\min N_j$ as the
addends should contain also a factor $(1-\fra1{3 N_j})$ to be exact:
for simplicity $O(1/N)$ corrections will be ignored (their inclusion
would imply trivial changes without affecting the physical
interpretation), \cite{Ga06c}.

An additive total derivative, $\dot R(\V X)$ in this case, of a
bounded quantity does not affect the long time fluctuations. Therefore
the average phase space contraction and the average entropy creation
rate have the {\it same average} $\s_+\=\e_+$ and, assuming
$\e_+\ne0$, the {\it same large deviations rate function} $\z(p)$ for
$p=\fra1{\s_+\t}\ig_0^\t \s(S_t (\V{{\dot X}},\V X))dt$ and for
$\fra1{\e_+\t}\ig_0^\t \e(S_t(\V{{\dot X}}, \V X))dt$.

Furthermore the equations of motion are {\it reversible} so that, under
the {\it Chaotic Hypothesis}, the {\it Fluctuation Theorem} yields,
\cite{GC95,Ga95b,Ru99,Ga06c,BGGZ05}, the symmetry property

\begin{equation}\z(-p)\,=\,\z(p)\,-\,p\,\e_+, \qquad |p|<p^*
\label{e1.4}\end{equation}
for the large deviations rate $\z(p)$ of $p$.  The identification
between phase space contraction and entropy creation rate is thus
motivated. It should be noted that Eq.(\ref{e1.3}) shows that even in
experiments, a case in which one hardly knows the equations of motion
and the phase space divergence, the divergence can be measured through
heat flow measurements, at least as far as its fluctuations over large
time are concerned, and give nontrivial consequences like the
fluctuation relation, Eq.(\ref{e1.4}).

\*
%\ifnum\tipo=2\pagina \fi
\0{\bf2. Quantum systems and thermostats}
\setcounter{section}{2}\setcounter{equation}{0}
\*

Is it possible to formulate a dissipation theory analogous to the one
developed for classical systems when the quantum nature of the system
in $\CC_0$ cannot be neglected?

At first it might seem almost impossible: in quantum systems average
kinetic energy is {\it not} identified with temperature; and all
motions are quasi periodic if the system is of finite size (as in our
examples in Sec.1), so that strictly speaking no chaos is possible.

A way out, explored in the literature, would be to imagine the
thermostats as infinite systems whose state far from $\CC_0$ is a
Gibbs state at a well defined temperature,
\cite{EPR99,Ku00,Ru00b,Ru01}. This is a point of view that could also
be taken in the classical case: however the recent progress in
classical statistical mechanics was sparked by the introduction of
{\it finite size} thermostat models and this is the path that will be
attempted here.

Thermostats have, usually, a macroscopic phenomenological nature:
in a way they should be regarded as classical macroscopic objects.
Therefore it seems natural to model them as such: thus their
temperature can be defined as the average kinetic energy and the
question of how to define it does not arise.

Consider the system in Fig.1 when the quantum nature of the particles
in $\CC_0$ cannot be neglected. Suppose first that the nonconservative force
$\V F(\V X_0)$ acting on the system vanishes, {\it i.e.} consider the
problem of heat flow through $\CC_0$.  Let $H$ be the operator on
$L_2(\CC_0^{3N_0})$, space of symmetric or antisymmetric wave
functions $\Ps$,

\begin{equation}
-\fra{\hbar^2}2\D_{\V X_0}+ U_0(\V X_0)+\sum_{j>0}\big(U_{0j}(\V X_0,\V
X_j)+U_j(\V X_j)\big)\label{e2.1}\end{equation}
where $\D_{\V X_0}$ is the Laplacian, and note that its spectrum
consists of eigenvalues $E_n=E_n(\{\V X_j\}_{j>0})$, for $\V X_j$
fixed.

A system--reservoirs model can be the {\it dynamical
system} on the space of the variables $\big(\Ps,(\{\V X_j\},\{\V{{\dot
X}}_j\})_{j>0}\big)$ defined by the equations (where
$\media{\cdot}_\Ps$ is the expectation in the state $\Ps$)

$$-i\hbar {\dot\Ps(\V X_0)}= \,(H\Ps)(\V X_0),\kern4mm{\rm and\ for}\ j>0 $$

\kern-6mm\begin{equation}\kern1mm\V{{\ddot X}}_j=-\Big(\dpr_j U_j(\V X_j)+
\media{\dpr_j U_j(\V X_0,\V X_j)}_\Ps\Big)-\a_j \V{{\dot X}}_j
\label{e2.2}\end{equation}

\kern-6mm$$\a_j\defi\fra{\media{W_j}_\Ps-\dot U_j}{2 K_j}, \qquad
W_j\defi -\V{{\dot X}}_j\cdot \V\dpr_j U_{0j}(\V X_0,\V
X_j)$$
here the first equation is Schr\"odinger's equation, the second is an
equation of motion for the thermostats particles, \cite{Ga06c},
similar to the one in Fig.1, whose notation for the particles labels
is adopted here too. Evolution maintains the thermostats
kinetic energies $K_j\=\fra12\V{{\dot X}}_j^2$ exactly constant so
that they can be used to define the thermostats temperatures $T_j$ via
$K_j=\fra32 k_B T_j N_j$, as in the classical case.

Let $\m_0(\{d\Ps\})$  be the {\it formal} measure on
$L_2(\CC_0^{3N_0})$ 

\begin{equation}
\Big(\prod_{\V X_0} d\Ps_r(\V X_0)\,d\Ps_i(\V X_0)
\Big)\,\d\Big(\ig_{\CC_0} |\Ps(\V Y)|^2\, d\V Y-1\Big)
\label{e2.3}\end{equation}
with $\Ps_r,\Ps_i$  real and imaginary parts of $\Ps$.
The formal phase space volume element $\m_0(\{d\Ps\})\times\n(d\V
X\,d\V{{\dot X}})$ with 

\begin{eqnarray}\n(d\V
X\,d\V{{\dot X}})\defi\prod_{j>0} \Big(d\V
X_j\,d\V{{\dot X}}_j\,\d(\V{{\dot X}}^2_j-3N_jk_B T_j)\Big)
\label{e2.4}\end{eqnarray}
is conserved, by the unitary property of the wave
functions evolution, just as in the classical case, {\it up
to the volume contraction in the thermostats}, \cite{Ga06c}. 

If $Q_j\defi\media{W_j}_\Ps$ and $R$ is as in Eq.(\ref{e1.3})
the contraction rate $\s$ of the volume element in Eq.(\ref{e2.4}) is
given by Eq.(\ref{e1.2}) with $\e$, that will be called {\it entropy
creation rate}, defined by Eq.(\ref{e1.3}).

In general solutions of Eq.(\ref{e2.2}) will not be quasi periodic
%\cite[Sec.3]{DGZ92},\cite{Go99}, 
and the Chaotic Hypothesis,
\cite{GC95b,Ga00}, can be assumed: if so the dynamics should select an
invariant distribution $\m$. The distribution $\m$ will give the
statistical properties of the stationary states reached starting the
motion in a thermostat configuration $(\V X_j,\V{{\dot X}}_j)_{j>0}$
randomly chosen with ``uniform distribution'' $\n$ on the spheres
$\V{{\dot X}}_j^2=3N_jk_B T_j$ and in a random eigenstate of $H$. The
distribution $\m$, if existing and unique, could be named the {\it SRB
distribution} corresponding to the chaotic motions of Eq.(\ref{e2.2}).

In the case of a system {\it interacting with a single thermostat} the
latter distribution should be equivalent to the canonical distribution. 

Hence an important consistency check for the model proposed in
Eq.(\ref{e2.2}) is that there should exist at least one stationary
distribution equivalent to the canonical distribution at the
appropriate temperature $T_1$ associated with the (constant) kinetic
energy of the thermostat: $K_1=\fra32 k_B T_1\,N_1$.  In the classical
case this is an established result, \cite{EM90},\cite{Ga00,Ga06c}.

Hence an important consistency check for the model proposed in
Eq.(\ref{e2.2}) is that there should exist at least one stationary
distribution $\m$ equivalent to the canonical distribution at the
appropriate temperature $T_1$ associated with the (constant) kinetic
energy of the thermostat: $K_1=\fra32 k_B T_1\,N_1$.  In the classical
case this is an established result, \cite{EM90},\cite{Ga00,Ga06c}.

The check should be performed also in the present case, thereby
providing further motivation and support for the model in Eq.(\ref{e2.2}).  A
first candidate for $\m$ might be to attribute a probability
proportional to $d\Ps\,d\V X_1\,d \dot{\V X}_1$ times

\begin{equation}
\sum_{n=1}^\io e^{-\b E_n}\d(\Ps-\Ps_n(\V
X_1)\,e^{i\f_n})\,{d\f_n}\,\d(\dot{\V X}_1^2-2K_1)\label{e2.5}\end{equation}
where $\Ps$ is the wave functions for the system in $\CC_0$ and ${\dot
X_1, X_1}$ are positions and velocities of the thermostat particles
and $\f_n\in [0,2\p]$ is a phase, $E_n=E_n(\V X_1)$ is the $n$-th
level of $H(\V X_1)$ with $\Ps_n(\V X_1)$ the corresponding
eigenfunction. The relation between the distribution in
Eq.(\ref{e2.5}) and EIG($\r$) in \cite{GLTZ05} should be
noted. However, as pointed out by a referee, Eq.(\ref{e2.5}) is not
invariant under the evolution Eq.(\ref{e2.2}) and it seems difficult
to exhibit explicitly an invariant distribution.  
\*

%\ifnum\tipo=2\pagina\fi
\0{\bf3. Some consequences. Conclusion.}
\setcounter{section}{3}\setcounter{equation}{0}
\*

The simplest case arises when $V\=0$: {\it i.e.} no nonconservative
force acts on $\CC_0$ and Eq.(\ref{e2.2}) models heat
flow, through $\CC_0$, between the various reservoirs.

Solutions of Eq.(\ref{e2.2}) are also {\it reversible}: time reversal
being the change in sign of the velocities and the conjugation of the
wave function $\Ps(\V X_0)$. Hence under the Chaotic Hypothesis the
{\it Fluctuation Theorem}, \cite{GC95},\cite{Ga00}, see
Eq.(\ref{e1.4}), would hold for the entropy creation rate fluctuations
in the SRB distribution: however since the phase space is infinite
dimensional corrections to Eq.(\ref{e1.4}) have to be expected for
large $p$, as in \cite{BGGZ05}, and should be discussed on a case by
case basis.  Note that the fluctuation theorem extension is also here
immediate once a model is properly formulated: as it was in the
analogous cases of infinite thermostats, \cite{Ku00}.

If a nonconservative force $\V F(\V X_0)$ acts on the system and has a
(multivalued) potential $V(\V X_0)$, so that $\V F(\V X_0)=-\V\dpr
V(\V X_0)$ is the force on the particles in $\CC_0$, then {\it in
absence of thermostats} generally the system will not reach a
stationary state: this is true both in the classical and in the
quantum cases. In the latter case the Schr\"odinger equation, with $H$
modified into $H+V$, will not have eigenvectors because of the
multivaluedness of the potential $V$: actually it will not even be
well defined. It should, however, be interpreted as an equation for
the wave function $\Ps$ defined on the ``covering space'' $\O$ of
$\CC_0^{3N_0}$ in which the potential $V(\V X_0)$ becomes single
valued, and the thermostats should have the effect of allowing
reaching a stationary state described by wave functions $\Ps_n(\V
X_0+\V\Bx(t))$ with $\Ps_n\in L_2(\O)$ and $\Bx(t)$ a suitable
``flow''.
\\
\hglue4mm Whether this really happens is, however, an open problem
even in the classical case; there it has been called the problem of
{\it efficiency} of the thermostats, \cite{Ga06c,GG07}, and it has been
studied only in a few numerical simulations involving long range or
short range particles interactions. {\it A fortiori} it is an open
problem in the quantum case.

Finally it should be stressed that Eq.(\ref{e2.2}) provides a model of
{\it finite} thermostat for a quantum system and therefore may be
suited for simulations and tests.

\* Identification of phase space contraction rate as the entropy
creation rate (up to an additive total time derivative) is an
achievement of the recent research in classical nonequilibrium
statistical mechanics, \cite{EM90}, which has led to the possibility
of nonequilibrium simulations without the need of considering infinite
thermostats and, subsequently, to general results like the Fluctuation
Theorems, \cite{GC95},\cite{Ga97,CG99}, and the possibility to test them
experimentally, \cite{BGGZ06},\cite{Ga06c,Ga06d}, and even to make use of
them in applications. In this note the proposal that rather
straightforward extensions to quantum nonequilibrium are possible has
been discussed.
\*

\0{\bf Acknowledgement:} I thank F. Zamponi for pointing out
reference \cite{Ku00}.
\*

%\*
%\pagina . \pagina 
\nota%\bibliography{0Bibcaos}

\begin{thebibliography}{18}
\expandafter\ifx\csname natexlab\endcsname\relax\def\natexlab#1{#1}\fi
\expandafter\ifx\csname bibnamefont\endcsname\relax
  \def\bibnamefont#1{#1}\fi
\expandafter\ifx\csname bibfnamefont\endcsname\relax
  \def\bibfnamefont#1{#1}\fi
\expandafter\ifx\csname citenamefont\endcsname\relax
  \def\citenamefont#1{#1}\fi
\expandafter\ifx\csname url\endcsname\relax
  \def\url#1{\texttt{#1}}\fi
\expandafter\ifx\csname urlprefix\endcsname\relax\def\urlprefix{URL }\fi
\providecommand{\bibinfo}[2]{#2}
\providecommand{\eprint}[2][]{\url{#2}}

\bibitem[{\citenamefont{Gallavotti}(2006{\natexlab{a}})}]{Ga06c}
\bibinfo{author}{\bibfnamefont{G.}~\bibnamefont{Gallavotti}},
  \bibinfo{journal}{Chaos} \textbf{\bibinfo{volume}{16}},
  \bibinfo{pages}{043114 (+6)} (\bibinfo{year}{2006}{\natexlab{a}}).

\bibitem[{\citenamefont{Gallavotti}(2000)}]{Ga00}
\bibinfo{author}{\bibfnamefont{G.}~\bibnamefont{Gallavotti}},
  \emph{\bibinfo{title}{Statistical Mechanics. A short treatise\\}}
  (\bibinfo{publisher}{Springer Verlag}, \bibinfo{address}{Berlin},
  \bibinfo{year}{2000}).

\bibitem[{\citenamefont{Gallavotti and Cohen}(1995{\natexlab{a}})}]{GC95}
\bibinfo{author}{\bibfnamefont{G.}~\bibnamefont{Gallavotti}} \bibnamefont{and}
  \bibinfo{author}{\bibfnamefont{E.~G.~D.} \bibnamefont{Cohen}},
  \bibinfo{journal}{Physical Review Letters} \textbf{\bibinfo{volume}{74}},
  \bibinfo{pages}{2694} (\bibinfo{year}{1995}{\natexlab{a}}).

\bibitem[{\citenamefont{Bonetto
  et~al.}(2006{\natexlab{a}})\citenamefont{Bonetto, Gallavotti, Giuliani, and
  Zamponi}}]{BGGZ05}
\bibinfo{author}{\bibfnamefont{F.}~\bibnamefont{Bonetto}},
  \bibinfo{author}{\bibfnamefont{G.}~\bibnamefont{Gallavotti}},
  \bibinfo{author}{\bibfnamefont{A.}~\bibnamefont{Giuliani}}, \bibnamefont{and}
  \bibinfo{author}{\bibfnamefont{F.}~\bibnamefont{Zamponi}},
  \bibinfo{journal}{Journal of Statistical Physics}
  \textbf{\bibinfo{volume}{123}}, \bibinfo{pages}{39}
  (\bibinfo{year}{2006}{\natexlab{a}}).

\bibitem[{\citenamefont{Gallavotti}(1995)}]{Ga95b}
\bibinfo{author}{\bibfnamefont{G.}~\bibnamefont{Gallavotti}},
  \bibinfo{journal}{Mathematical Physics Electronic Journal (MPEJ)}
  \textbf{\bibinfo{volume}{1}}, \bibinfo{pages}{1} (\bibinfo{year}{1995}).

\bibitem[{\citenamefont{Ruelle}(1999)}]{Ru99}
\bibinfo{author}{\bibfnamefont{D.}~\bibnamefont{Ruelle}},
  \bibinfo{journal}{Journal of Statistical Physics}
  \textbf{\bibinfo{volume}{95}}, \bibinfo{pages}{393} (\bibinfo{year}{1999}).

\bibitem[{\citenamefont{Eckmann et~al.}(1999)\citenamefont{Eckmann, Pillet, and
  Bellet}}]{EPR99}
\bibinfo{author}{\bibfnamefont{J.~P.} \bibnamefont{Eckmann}},
  \bibinfo{author}{\bibfnamefont{C.~A.} \bibnamefont{Pillet}},
  \bibnamefont{and} \bibinfo{author}{\bibfnamefont{L.~R.}
  \bibnamefont{Bellet}}, \bibinfo{journal}{Communications in Mathematical
  Physics} \textbf{\bibinfo{volume}{201}}, \bibinfo{pages}{657}
  (\bibinfo{year}{1999}).

\bibitem[{\citenamefont{Kurchan}(2000)}]{Ku00}
\bibinfo{author}{\bibfnamefont{J.}~\bibnamefont{Kurchan}},
  \bibinfo{journal}{cond-mat/0007360}  (\bibinfo{year}{2000}).

\bibitem[{\citenamefont{Ruelle}(2000)}]{Ru00b}
\bibinfo{author}{\bibfnamefont{D.}~\bibnamefont{Ruelle}},
  \bibinfo{journal}{Journal of Statistical Physics}
  \textbf{\bibinfo{volume}{98}}, \bibinfo{pages}{55} (\bibinfo{year}{2000}).

\bibitem[{\citenamefont{Ruelle}(2001)}]{Ru01}
\bibinfo{author}{\bibfnamefont{D.}~\bibnamefont{Ruelle}},
  \bibinfo{journal}{Communications in Mathematical Physics}
  \textbf{\bibinfo{volume}{224}}, \bibinfo{pages}{3} (\bibinfo{year}{2001}).

\bibitem[{\citenamefont{Gallavotti and Cohen}(1995{\natexlab{b}})}]{GC95b}
\bibinfo{author}{\bibfnamefont{G.}~\bibnamefont{Gallavotti}} \bibnamefont{and}
  \bibinfo{author}{\bibfnamefont{E.~G.~D.} \bibnamefont{Cohen}},
  \bibinfo{journal}{Journal of Statistical Physics}
  \textbf{\bibinfo{volume}{80}}, \bibinfo{pages}{931}
  (\bibinfo{year}{1995}{\natexlab{b}}).

\bibitem[{\citenamefont{Evans and Morriss}(1990)}]{EM90}
\bibinfo{author}{\bibfnamefont{D.~J.} \bibnamefont{Evans}} \bibnamefont{and}
  \bibinfo{author}{\bibfnamefont{G.~P.} \bibnamefont{Morriss}},
  \emph{\bibinfo{title}{Statistical Mechanics of Non{\-}equilibrium Fluids}}
  (\bibinfo{publisher}{Academic Press}, \bibinfo{year}{1990}).

\bibitem[{\citenamefont{Goldstein et~al.}(2006)\citenamefont{Goldstein,
  Lebowitz, Tumulka, and N.Zangh\`\i}}]{GLTZ05}
\bibinfo{author}{\bibfnamefont{S.}~\bibnamefont{Goldstein}},
  \bibinfo{author}{\bibfnamefont{J.}~\bibnamefont{Lebowitz}},
  \bibinfo{author}{\bibfnamefont{R.}~\bibnamefont{Tumulka}}, \bibnamefont{and}
  \bibinfo{author}{\bibnamefont{N.Zangh\`\i}}, \bibinfo{journal}{Journal of
  Statistical Physics} \textbf{\bibinfo{volume}{125}}, \bibinfo{pages}{1193}
  (\bibinfo{year}{2006}).

\bibitem[{\citenamefont{Garrido and Gallavotti}(2007)}]{GG07}
\bibinfo{author}{\bibfnamefont{P.}~\bibnamefont{Garrido}} \bibnamefont{and}
  \bibinfo{author}{\bibfnamefont{G.}~\bibnamefont{Gallavotti}},
  \bibinfo{journal}{Journal of Statistical Physics}
  \textbf{\bibinfo{volume}{cond-mat/0607231 and mp$\_$arc 06-197}},
  \bibinfo{pages}{doi: 10.1007/s10955} (\bibinfo{year}{2007}).

\bibitem[{\citenamefont{Gallavotti}(1999 and chao-dyn/9703007)}]{Ga97}
\bibinfo{author}{\bibfnamefont{G.}~\bibnamefont{Gallavotti}},
  \bibinfo{journal}{Annales de l' Institut H. Poincar\'e}
  \textbf{\bibinfo{volume}{70}}, \bibinfo{pages}{429} (\bibinfo{year}{1999 and
  chao-dyn/9703007}).

\bibitem[{\citenamefont{Cohen and Gallavotti}(1999)}]{CG99}
\bibinfo{author}{\bibfnamefont{E.~G.~D.} \bibnamefont{Cohen}} \bibnamefont{and}
  \bibinfo{author}{\bibfnamefont{G.}~\bibnamefont{Gallavotti}},
  \bibinfo{journal}{Journal of Statistical Physics}
  \textbf{\bibinfo{volume}{96}}, \bibinfo{pages}{1343} (\bibinfo{year}{1999}).

\bibitem[{\citenamefont{Bonetto
  et~al.}(2006{\natexlab{b}})\citenamefont{Bonetto, Gallavotti, Giuliani, and
  Zamponi}}]{BGGZ06}
\bibinfo{author}{\bibfnamefont{F.}~\bibnamefont{Bonetto}},
  \bibinfo{author}{\bibfnamefont{G.}~\bibnamefont{Gallavotti}},
  \bibinfo{author}{\bibfnamefont{A.}~\bibnamefont{Giuliani}}, \bibnamefont{and}
  \bibinfo{author}{\bibfnamefont{F.}~\bibnamefont{Zamponi}},
  \bibinfo{journal}{Journal of Statistical Mechanics} p.
  \bibinfo{pages}{P05009} (\bibinfo{year}{2006}{\natexlab{b}}).

\bibitem[{\citenamefont{Gallavotti}(2006{\natexlab{b}})}]{Ga06d}
\bibinfo{author}{\bibfnamefont{G.}~\bibnamefont{Gallavotti}},
  \bibinfo{journal}{Journal of Statistical Mechanics (JSTAT)} pp.
  \bibinfo{pages}{P10011 (+9)} (\bibinfo{year}{2006}{\natexlab{b}}).

\end{thebibliography}
\bibliographystyle{apsrev}

\ifnum\tipo=2
\0e-mail: {\tt giovanni.gallavotti@roma1.infn.it}
\\
web: {\tt http://ipparco.roma1.infn.it}
\hfill
\revtex
\fi

\ifnum\tipo=1
\0e-mail: {\tt giovanni.gallavotti@roma1.infn.it},
\\
web: {\tt http://ipparco.roma1.infn.it}
\hfill\revtex
\fi
\end{document}